# Detection of field-free magnetization switching through thermoelectric effect in Ta/Pt/Co/Pt with significant spin-orbit torque and competing spin currents

*Raghvendra Posti, Abhishek Kumar, Dhananjay Tiwari, and Debangsu Roy[*]*

[1]Department of Physics, Indian Institute of Technology Ropar, Rupnagar 140001, India

[2] Advanced safety & User Experience, Aptiv Services, Krakow, Poland

**ABSTRACT:** Application of sufficient lateral current to a heavy metal (HM) can switch the perpendicular magnetization orientation of adjacent ferromagnetic layer (FM) through spin-orbit torques (SOTs). The choice of the HM and its arrangement plays a major role for the SOT induced magnetization switching in magnetic heterostructures. Here, in asymmetric Pt/Co/Pt heterostructures, anti-damping (AD) SOT prevails. Ta addition to this stack (Ta/Pt/Co/Pt) give rise to several compelling effects viz. competing spin currents (due to opposite spin-Hall angles of adjacent Ta and Pt layers), significant AD-SOT, thermoelectric effects (particularly, anomalous Nernst effect (ANE)), and enhanced perpendicular magnetic anisotropy. For this Ta/Pt/Co/Pt stack, the AD-SOT values are stabilized to that of the Pt/Co/Pt stack, which is significant than what is expected for a stack with competing spin currents. Current-induced field-free magnetization switching was absent in uniformly grown

Ta/Pt/Co/Pt stack. It was observed that a thickness gradient is essential to assist the field-free magnetization switching in these heterostructures. Further, the thermoelectric effects are utilized to develop a technique to detect the field-free magnetization switching. This technique detects the second harmonic ANE signal as a reading mechanism. Using ANE symmetry with the applied current, the switching can be detected in a single current sweep which was corroborated to the conventional DC Hall method.

1. **Introduction**

Recent years have witnessed a renewed interest in developing spin-based logic and memory devices capable of operating at high frequency with low energy consumption [1, 2, 3]. However, spin-transfer torque (STT) based magnetic random access memory devices experiences low endurance issues due to the flow of high current density through the tunnel barrier during writing operation [4]. In this regard, MRAM utilizing spin-orbit torque (SOT) mechanism for magnetization switching becomes of paramount importance owing to better endurance, faster access time and lower energy consumption in comparison to STT-MRAM. In SOT-MRAM, lateral current flowing through the heavy metal (HM) layer generates spin current due to the bulk spin-Hall effect (SHE) [3, 5] or/and interfacial Rashba-Edelstein effect (REE) [6]. The spin current in-turn induces a torque in the adjacent ferromagnetic (FM) layer leading to the magnetization reversal in the FM layer in HM/FM heterostructures. Here, the spin-torques generally comprised of two orthogonal components namely anti-damping like torque (AD-SOT) and field-like torque (FL-SOT) [3, 7]. Moreover, for realizing high density SOT-MRAM, the robust perpendicular magnetic anisotropy (PMA) in the HM/FM heterostructures is desired to maintain a significant energy barrier prohibiting the thermal fluctuation.

SHE induced AD-SOT is the dominating torque which can manipulate magnetization orientation in PMA based magnetic heterostructures [1, 3, 8]. In order to exploit the nature of SHE and engineer the SOT in magnetic heterostructures, different stacks of HMs and FMs possessing PMA are used, unlike the conventional HM/FM/oxide stacks [1, 8]. Amongst them, HM/FM/HM stacks with PMA are of particular interest where the HM layers can be of dissimilar materials [9-12]. By carefully engineering the HM layer according to their spin-Hall angles ($\theta_{SH}$), SOTs and other magnetic properties in HM/FM/HM heterostructures become tuneable. For instance, Pt/Co/Ta enhances SOT efficiency [12] as both the HMs have opposite $\theta_{SH}$, whereas, a symmetric thickness of Pt in the Pt/Co/Pt stack diminishes the SOT efficiency [10, 11]. Remarkably, the Ta/Pt/Co/Ta stack enhances the SOT efficiency as well as the anisotropy field [9]. Thus, HM/FM/HM stack with PMA and varying spin current is considered as an ideal composite FM layer for investigating different SOT studies. The main motivation behind controlling these SOTs is to understand and achieve magnetization switching without the application of an external magnetic field.

In the last decade, various mechanisms have been proposed to achieve magnetization switching without the application of any external symmetry-breaking magnetic field, such as the use of a wedge layer (either FM [13-15], HM [16, 17], or oxide [18]), exchange bias field of antiferromagnets [19] and so on. In a recent study, competing spin currents from two successive HM layers having opposite signs of $\theta_{SH}$ were also reported to show a field-free switching [20] while others have argued wedge deposition of the HM layer acts as a source of field-free switching [14, 17]. In these studies, the case of vanishing AD-SOT was extensively investigated whereas the case of competing spin currents with considerable AD-SOT is still lacking.

In the present study, we had characterized the SOTs in Ta/Pt/Co/Pt stacks in comparison to the Pt/Co/Pt asymmetric stack. Here, Pt/Co/Pt asymmetric stack was considered as a model system for studying different SOT behaviour. Further, we had introduced a Ta underlayer to the Pt/Co/Pt asymmetric stack to investigate the emergence of various physical phenomena due to the addition of the Ta layer and studied its implication in SOT behaviour and field-free switching. We had used low field AC harmonic Hall voltage technique [21, 22] and magnetization angle-resolved harmonic Hall technique [23, 24] for estimating the SOT contribution in Pt/Co/Pt and Ta/Pt/Co/Pt stack, respectively. In Ta/Pt/Co/Pt stack, significant thermoelectric effects persist due to the differences in the resistivity and thermal conductivity of the layers [25, 26] in comparison to Pt/Co/Pt. It was found that AD-SOT contribution in Ta/Pt/Co/Pt stack is similar to the value of AD-SOT for Pt/Co/Pt stack. We observed field-free switching in Ta/Pt/Co/Pt stack using conventional DC Hall methodology. Notably, in this stack, the thermoelectric effect, competing spin currents as well as considerable AD-SOT contribution coexist. It is noteworthy, that the combination of competing spin currents and non-zero value of AD-SOT in Ta/Pt/Co/Pt did not yield any field-free switching in the uniformly grown Ta/Pt/Co/Pt stack. The field-free switching in Ta/Pt/Co/Pt stack was observed when the Co layer had a thickness gradient. This implies that the thickness gradient induced tilted anisotropy is the origin of field-free switching in our thin-film stacks. This finding was further confirmed with the micromagnetic simulation. Moreover, we had utilized the thermoelectric effects in Ta/Pt/Co/Pt as a reading mechanism for magnetization switching. We developed an AC Hall resistance reading technique using anomalous Nernst effect (ANE) to detect the field-free switching in Ta/Pt/Co/Pt apart from estimating the switching current values.

## 2. EXPERIMENTAL DETAILS

The thin film stacks of Pt(3)/Co(0.6)/Pt(6) (hereafter denoted as Pt stack) and Ta(3)/Pt(3)/Co(0.6)/Pt(6) (hereafter denoted as Ta stack) were deposited onto thermally oxidized Si/SiO$_2$ substrate by dc-magnetron sputtering. Here, Pt(3) & Ta (3) were the bottom layer for both the Pt & Ta stack, respectively. The thickness of the films indicated in the parenthesis are in nanometers. All films were deposited at room temperature after achieving a base pressure ~10$^{-8}$ Torr at a working pressure of 3 mTorr. The Co wedge layer had a thickness gradient of 0.06 nm/cm. Six terminal hall bar devices (Figure 1(a)) with a dimension of $135 \mu m \times 12 \mu m$ were patterned using standard photo-lithograph and plasma etching process for the subsequent transport measurements. In our Hall bar devices, current channels along and perpendicular to the Co thickness gradient direction, were fabricated.

Notably, devices with current channel parallel to the wedge direction were considered to have insignificant thickness gradient along y-direction (see Figure 1(a) for the coordinates) and considered as uniformly grown Co devices. Unpatterned films in both the stacks (Pt stack & Ta stack) were utilized for the magnetization measurements using vibration sample magnetometry (VSM). Figure 1(b) shows the variation of the magnetization of the stacks as a function of applied field along *z*-direction ($H_z$). The square shaped hysteresis loop as depicted in Figure 1(b) confirms the PMA behaviour in both the stacks. The saturation magnetization ($M_s$) was found to be ~1057 emu/cc for Pt stack and ~1135 emu/cc for Ta stack, respectively. The occurrence of the PMA in both the stack was further corroborated by performing anomalous Hall effect (AHE) resistance measurement as a function of $H_z$ which is depicted in the inset of Figure1(b). The square shaped hysteresis behaviour of AHE measurement confirms the PMA in both Pt stack and Ta stack.

For the quantitative study of SOTs, AC measurement techniques utilizing lock-in amplifiers (EG&G 7265) with reference frequency of 577.13 Hz were used for both Pt and Ta stacks. Current induced magnetization switching was probed using the DC Hall measurement technique (utilizing Keithley 2450 source meter and Keithley 2182A nanovoltmeter) and $2^{nd}$ harmonic AC Hall technique. All transport measurements were carried out at room temperature.

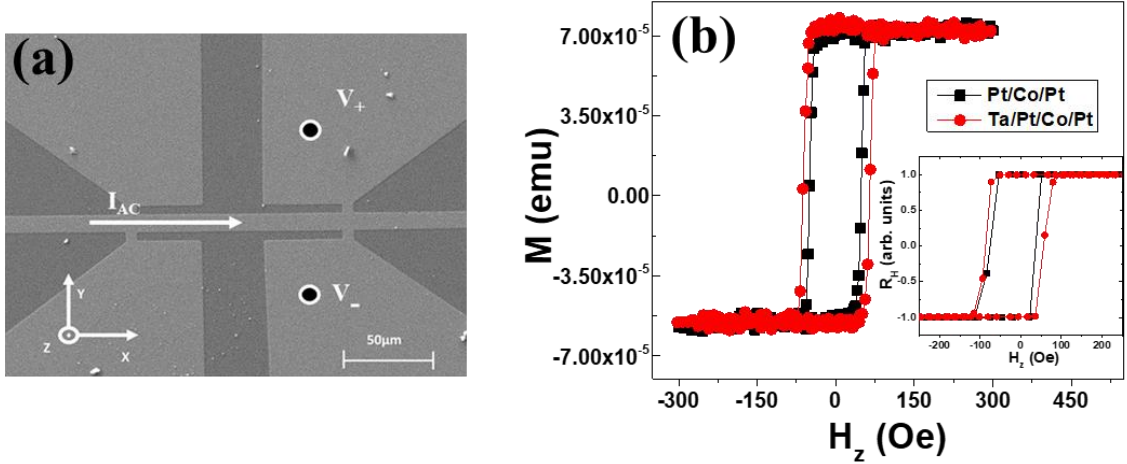

**FIG. 1(a)** Scanning electron microscopy image of the six-terminal Hall bar devices with Hall measurement geometry. **(b)** Magnetization measurements by VSM as a function of out-of-plane applied field ($H_z$) and inset shows AHE resistance measurements vs. $H_z$.

## 3. RESULTS AND DISCUSSION

### 3.1. SOT measurement for the Pt stack

For Pt stack, effective fields induced by AD-SOT ($H_{AD}$) and FL-SOT ($H_{FL}$) were extracted using low field harmonic Hall measurement method. The results of harmonic Hall voltage measurements with varying external magnetic field for Pt stack are shown in Figure 2. Figure 2(a) depicts the experimental geometry and coordinate system along with the material stack. Figure 2(b) shows the representative plots for the in-phase first harmonic Hall voltage $V_H^{\omega}$ (left y-axis) and out-of-phase second harmonic voltage $V_H^{2\omega}$ (right y-axis) as a function of the external magnetic field applied in the direction of current ($H_x$). The variation of $V_H^{\omega}$ (left y-axis) and $V_H^{2\omega}$ (right y-axis) as a function of the external magnetic field transverse to the current ($H_y$) is shown in Figure 2(c). A small out-of-plane field was applied during the experiment to prevent the domain nucleation in the low field range. The effective fields $H_{AD}$, and $H_{FL}$ generated by current-induced AD-SOT and FL-SOT are calculated using [21]:

$$H_{AD(FL)} = \frac{B_{AD(FL)} \pm 2\xi B_{FL(AD)}}{1-4\xi^2} \qquad (1)$$

where $B_{AD(FL)} = -2\left(\frac{\partial V_H^{2\omega}}{\partial H_{X(Y)}}\right)\bigg/\left(\frac{\partial^2 V_H^{\omega}}{\partial H^2_{X(Y)}}\right)$ and $\xi = \Delta R_{PHE}/\Delta R_{AHE}$ is the ratio of planar Hall (PHE) and anomalous Hall resistances (see supporting information for AHE and PHE measurements). Here $\pm$ sign corresponds to up-and down-magnetization, respectively. Further, the variation of $V_H^{\omega}$ and $V_H^{2\omega}$ with the applied magnetic field using Eq. (1) leads to the estimation of the $H_{AD}$, and $H_{FL}$ values for Pt stack.

The variation of the extracted $H_{AD}$, and $H_{FL}$ values under different applied currents are depicted in Figure 2(d) and Figure 2(e), respectively. $H_{AD}$ values increases with the applied current at a rate of $+1.6$ Oe/mA ($-1.8$ Oe/mA) for up-magnetization (down-magnetization) whereas $H_{FL}$ in Pt stack shows no particular trend and is always an order smaller than $H_{AD}$. The similar interfaces of Pt on both sides of the Co layer show the insignificant effect of Rashba-Edelstein effect (REE) and, as an outcome, FL-SOT is negligible compared to the AD-SOT [10]. The significant $H_{AD}$ values suggest that, due to the Pt thickness asymmetry on both side of Co thin film, a net spin current flows from the thicker Pt in the Pt stack. The intuitive picture is depicted in Figure 2(a) where a net spin current ($J_s^{efffective}$) acts upon the Co thin film which is the resultant of the countering contribution from the thicker Pt on top ($Pt_{top}$) & thinner Pt on bottom ($Pt_{bottom}$) of the Co thin film.

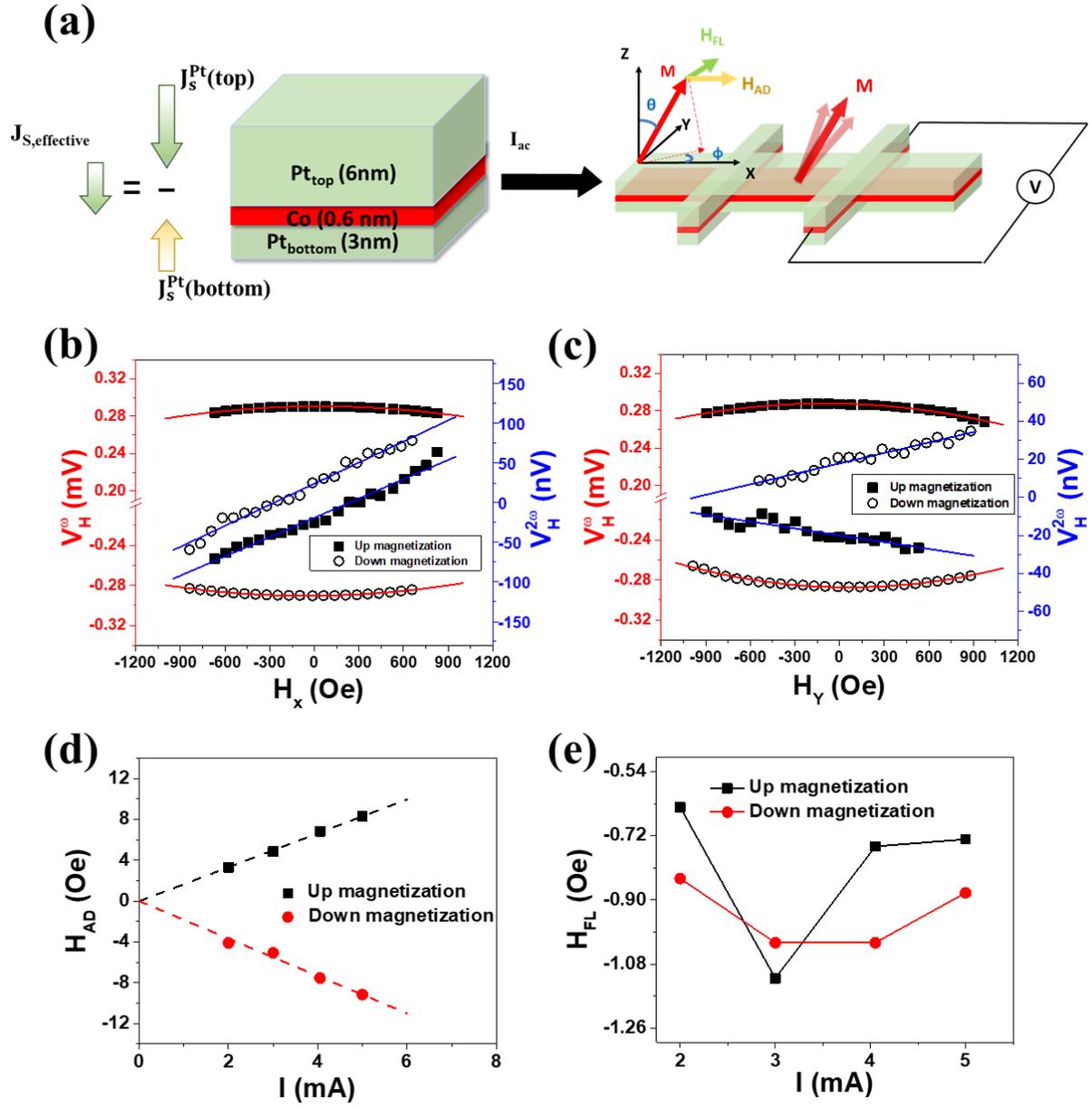

**FIG. 2(a)** Pt stack with directions of spin currents ($J_S$) flowing from each HM layer and an effective spin current direction, along with the schematic diagram for the AC Hall measurement with its coordinate geometry. **(b)** First harmonic ($V_H^\omega$) and second harmonic ($V_H^{2\omega}$) voltage vs. magnetic field sweep in *x*-direction ($H_x$) under ~4 mA current. **(c)** $V_H^\omega$ and $V_H^{2\omega}$ vs. magnetic field sweep in *y*-direction ($H_y$) under ~4 mA current. **(d)** $H_{AD}$ and **(e)** $H_{FL}$ as a function of applied AC currents.

### 3.2. Pt stack and Ta stack comparison

Next, we compared the second harmonic Hall resistance ($R_H^{2\omega}$) as a function of magnetic field, swept along the current direction ($H_x$) for both Pt and Ta stacks which is depicted in Figure 3. The $R_H^{2\omega}$ vs $H_x$ curve for Pt stack (Figure 3(a)) differs distinctly from the $R_H^{2\omega}$ vs. $H_x$ plot of Ta stack (Figure 3(b)). The Pt stack curve shows a generic AD-SOT behaviour with asymmetric $R_H^{2\omega}$ about the field axis (as, $H_{AD} \propto m \times y$)[23, 27]. In case of Ta stack, the $R_H^{2\omega}$ vs. $H_x$ plot exhibits a significant hysteric nature in low field regime. This is attributed to the presence of thermoelectric effects [24]. The electrical current passing through the sample leads to Joule heating which in turn creates the temperature gradient ($\nabla T$) due to mismatch in the resistivity and thermal conductivity in the stack after introducing the Ta layer [25, 26]. Notably, the temperature gradient in Ta stack can generate anomalous Nernst effect (ANE) along with longitudinal spin Seebeck effect (SSE) [26, 28]. However, the contribution of the SSE is negligible since the Ta stack possesses significant PMA [28]. Thus, the hysteric nature is attributed to the presence of ANE in Ta stack. Note that, for Pt stack, the Pt (3 nm) and Pt (6 nm) undergoes different growth conditions (initial Pt(3) was grown on Si/SiO2 wafer whereas Pt(6) was grown on Co layer) which may lead to the difference in the resistivities of the Pt thin films (top & bottom) in comparison to the Co thin film. This could lead to the ANE in Pt stack as well. Moreover, while analyzed closely, the inset of Figure 3(a) shows very faint hysteric signal. In order to comprehend the relative contribution of the ANE in Pt stack & Ta stack, we have measured the variation of the $R_H^{2\omega}$ with the external field ($H_z$) swept along the easy axis i.e., the *z*-axis of the sample [28]. Figure 3(c) exhibits the variation of the $R_H^{2\omega}$ vs $H_z$ for both Pt stack & Ta stack. The occurrence of the feeble hysteresis for Pt stack in comparison to the Ta stack confirms that the ANE signal is dominant (negligible) in Ta stack (Pt stack). Notably, the voltage generated due to ANE is proportional to $m \times \nabla T$, (*m* being the magnetization of the thin film) which has similar magnetization dependency as $H_{AD}$. In this regard, while estimating $H_{AD}$ through the measurement of $R_H^{2\omega}$, the contribution of the ANE needs to be separated else it would lead to the overestimation of the $H_{AD}$ values in Ta stack.

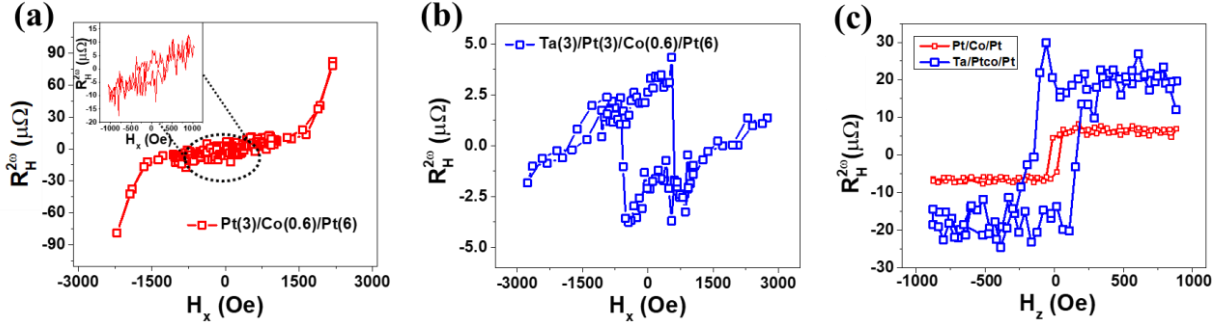

**FIG. 3(a)** $R_H^{2\omega}$ vs $H_x$ for Pt stack. Inset: zoom-in view of $R_H^{2\omega}$ vs $H_x$ in lower field regime. **(b)** $R_H^{2\omega}$ vs $H_x$ for Ta stack, **(c)** $R_H^{2\omega}$ vs $H_z$ comparison for Pt stack and Ta, measurements are performed at current density $J_{AC} \sim 1 \times 10^{10}$ A/m$^2$.

### 3.3. SOT measurement for the Ta stack

To extricate the AD-SOT generated effective field from the ANE contribution, we followed a low field second harmonic magnetization rotation technique recently developed by H Yang *et al*. [24]. This method avoids the application of high magnetic field ($H_{ext} > H_k$) while measuring the thermoelectric effects along with the SOT fields. In this method, the $R_H^{\omega}$ and $R_H^{2\omega}$ were measured to quantify AD-SOT and ANE contribution by rotating the sample in *xz*-plane in presence of a constant magnetic fields ($H_{ext}$) [23, 24] (experimental geometry shown in Figure 4(a)). The contribution of FL-SOT is negligible for Ta stack due to similar interfaces about Co layer (see supporting information for details). Considering trifling contribution from planar Hall effect and FL-SOT for Ta stack, the magnetization angle dependent ($\theta$) $R_H^{2\omega}$ in presence of $H_{ext}$ can be expressed as [24]:

$$R_H^{2\omega} = \left(\frac{\Delta R_{AHE}}{2}\frac{H_{AD}}{H_{ext}+H_k} + I_0 \alpha \nabla T\right) sin\theta. \qquad (2)$$

where $\alpha$ is the ANE coefficient and $I_0$ is the applied AC current defined by $I_{AC} = I_0 sin(\omega t)$. The magnetization angle $\theta$, was found using $\theta = cos^{-1}(Norm. R_H^\omega)$ where the normalized $R_H^\omega$ was obtained as a function of magnetic field angle ($\theta_H$) (Figure 4(b)). Equation (1) can further be modified as:

$$\frac{dR_H^{2\omega}}{d(sin\theta)} = \frac{\Delta R_{AHE}}{2}\frac{H_{AD}}{H_{ext}+H_k} + I_0\alpha\nabla T \qquad (3)$$

Thus, it is important to plot the variation of $R_H^{2\omega}$ vs. $\theta$ which is depicted in Figure 4(c). According to the Eq. (3) the slope of the variation of $R_H^{2\omega}$ vs. $sin\theta$ would give rise to the quantification of the ANE voltage along with AD-SOT. Figure 4(d) shows the plot of $\frac{dR_H^{2\omega}}{d(sin\theta)}$ vs. $\frac{1}{H_{ext}+H_k}$. The slope of this graph leads to the estimation of the AD-SOT whereas the intercept predicts the ANE contributions in Ta stack. In our calculation we have used $H_k$ ~1.4T for Ta stack (see supporting information for the anisotropy field calculation). From Figure 4(d) and using Eq. (3) we have obtained $H_{AD}$~7 Oe and $V_{ANE}$=0.32 $\mu V$ [23, 24].

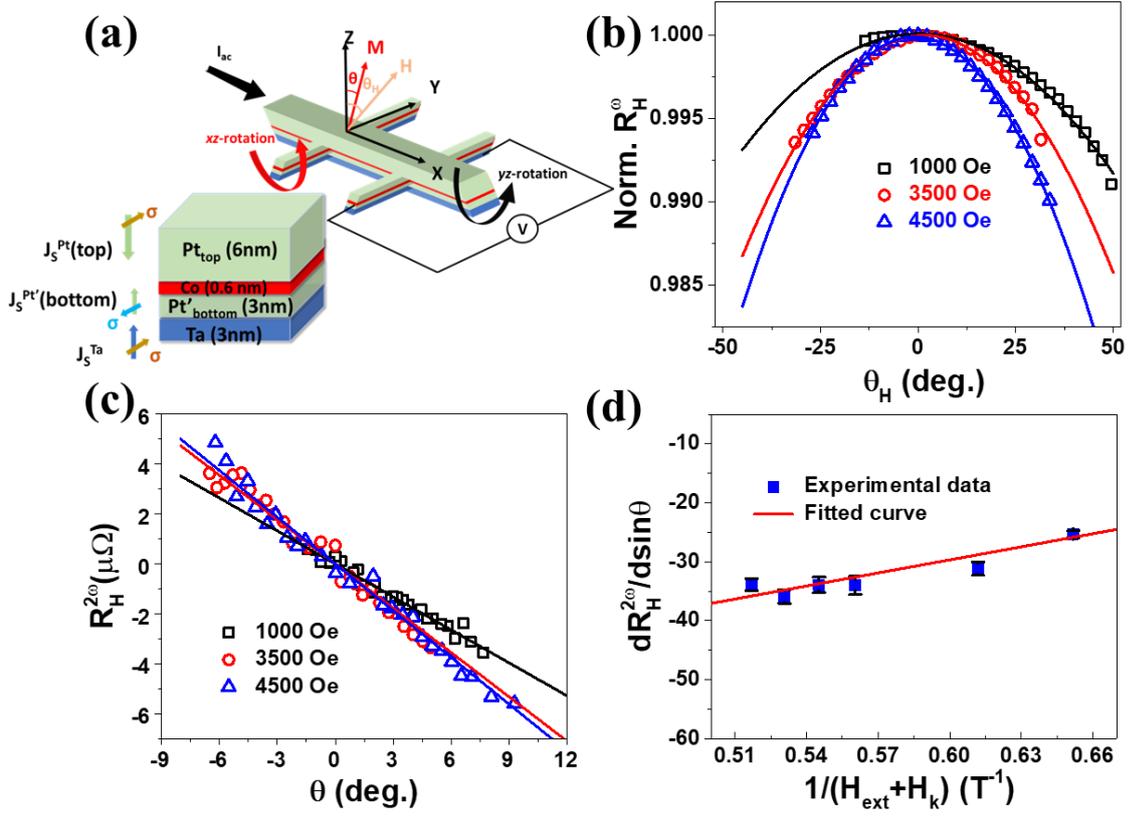

**FIG. 4(a)** Illustration of the experimental geometry and spin current ($J_S$) and polarization directions ($\sigma$) for each HM layer in Ta/Pt/Co/Pt stack. **(b)** Normalized $R_H^\omega$ vs. $\theta_H$. **(c)** $R_H^{2\omega}$ vs. $\theta$ at different external magnetic fields. **(d)** Contributions of SOT and ANE as a function of $1/(H_{ext}+H_k)$. Measurements are performed at $I_{AC}= 4.4$ mA.

Remarkably, $H_{AD}$ for Ta stack matches well with the value of $H_{AD}$ for Pt stack when measured with the same current density of ~ $3\times10^{10}$ A/m$^2$. The intuitive picture leading to this similarity in the $H_{AD}$ values in both the samples is depicted in Figure 4(a). In Pt stack, AD-SOT was generated because of thickness asymmetry in Pt$_{top}$ and Pt$_{bottom}$ (as shown in Figure 4(a)). The effective spin current acting on the Co layer for Pt stack can be expressed as $J_s^{Pt}$ (top) $- J_s^{Pt}$ (bottom). Similarly, for Ta stack, the resultant spin current can be expressed as $J_s^{Pt}$ (top) $- J_s^{Pt'}$ (bottom) $+ J_s^{Ta}$ (bottom). Notably, the addition of Ta underlayer produces a spin current that has similar spin polarization as Pt$_{top}$ since Pt and Ta has opposite $\theta_{SH}$. One has to note that the Pt$_{bottom}$ layer in Pt stack is directly grown on Si/SiO$_2$ substrate whereas in Ta stack (Pt′$_{bottom}$) it is grown on Ta. It is expected that the quality of the Pt$_{bottom}$ layer in Pt stack is inferior to the Pt′$_{bottom}$ layer in Ta stack. Note that, though 3nm Pt layer separates the Ta layer from Co layer in Ta stack, a less dominant spin current originating in Ta layer can still pass-through the Pt layer and acts on Co. This scenario is equivalent to the situation reported in Ref [17] where the W/Pt/Co/Pt shows a gradual decrease of AD-SOT efficiency (from W dominating to Pt dominating) as the thickness of Pt increases. Notably, In Ref [17] the Pt thickness was always greater than the spin diffusion length (~1.2nm). This in turn indicates that $J_s^{Pt'}(bottom) - J_s^{Ta}(bottom)$ (for Ta stack) $\approx J_s^{Pt}(bottom)$ (for Pt stack). This results in the comparable value of the effective spin current acting on the Co layer for both the stack leading to the similar value of H$_{AD}$ for both the Pt and Ta stack, respectively.

### 3.4. SOT induced magnetization switching

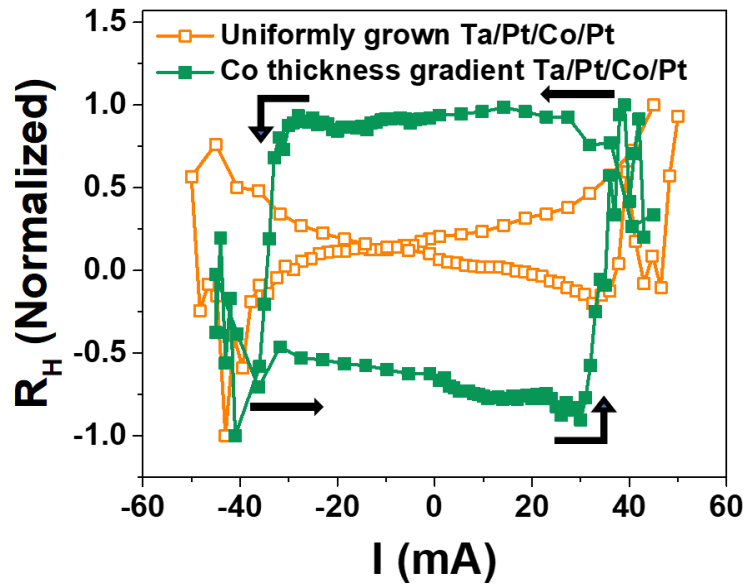

**Fig. 5** R$_H$ vs. I for Co wedge and uniformly grown Ta stacks.

The presence of considerable AD-SOT in Ta stack was further confirmed by performing magnetization switching experiments through DC Hall resistance ($R_H$) measurements in presence of a symmetry breaking in-plane magnetic field (See supporting information for further details). Further, we had achieved the field-free magnetization switching by measuring $R_H$ as a function of DC current pulses. The typical field-free switching behaviour of the Ta stack is demonstrated in Figure 5. Notably, the field-free switching was achieved for the sample with Co deposited in the wedge form unlike the uniformly deposited Co sample. Until now, the field-free magnetization switching experiments in competing current-based heterostructures have been executed for negligible AD-SOT values. However, in our case, we probed the magnetization switching in Ta stack with competing spin currents and significant AD-SOT. It is noteworthy, that the field-free magnetization switching was not achieved in the uniformly grown Ta stack confirming that field-free switching cannot be accomplished by competing spin currents alone.

The thickness gradient of FM tilts the out-of-plane anisotropy with a very small angle [13] which helps to achieve the field-free magnetization switching. It was further confirmed by the micromagnetic simulation. Micromagnetic simulations for Ta stack were performed using OOMMF [29]. In order to replicate the current-induced SOT switching experiments, we have used simulation parameters found from the experiments and literatures. Input parameters include sample dimension of $100 \times 100$ nm$^2$, saturation magnetization $M_s$ =1135 emu/cc, exchange stiffness constant [30] A=2×10$^{-11}$ J/m, DMI constant [31] D=0.2 mJ/m$^2$, $H_k$=1.4 T, spin-Hall angle $\theta_{SH}$= 0.49 and, critical current density J=3×10$^{11}$ A/m$^2$, respectively. In this simulation, we have probed the time evolution of the out-of-plane component of the magnetization ($m_z$) after applying a current pulse. Figure 6(a) depicts the time evolution of $m_z$ for the uniformly grown Ta stack with the applied current pulse (shaded region) of density $3\times10^{11}$ A/m$^2$ for duration of 4 ns in presence of an in-plane symmetry breaking field of 100 Oe. The $m_z$ found to switch due to the applied current pulse. However, in absence of the symmetry breaking field, the magnetization does not switch as evident from the simulation results presented in Figure 6(b) for the uniformly grown Ta stack. Thus, the current pulse alone was unable to switch the magnetisation for the uniformly grown Ta stack. Further, Figure 6(c) exhibits the scenario when the applied current pulse has lesser magnitude compared to the critical current density of $3\times10^{11}$ A/m$^2$ required to induce the switching in uniformly grown Ta stack in presence of 100 Oe symmetry breaking field. In this scenario, the magnetization did not switch its orientation. These results validate our experimental studies on the absence of field-free magnetization switching in uniform Ta stack. To mimic the experimental geometry of the Ta stack with Co wedge, we had considered a small tilt (~5°) in the anisotropy direction and we had performed the micromagnetic simulation in presence of current pulses without any symmetry breaking magnetic field. Figure 6(d) shows the time evolution of $m_z$ in presence of a current pulse and it is found that the magnetization indeed switches its orientation without the application of an external field. This corroborates our earlier experimental findings of the field-free magnetization switching due to the Co thickness gradient induced tilted anisotropy with competing spin currents and considerable AD-SOT. Further, the threshold value of the current pulse density of $3\times10^{11}$ A/m$^2$ found in micromagnetic simulation matches well with the experimentally obtained current pulse density from field free switching behaviour in the Ta stack with Co wedge. All these simulation results confirm that the thickness gradient induced tilted anisotropy helps to achieve the field-free switching of the magnetization.

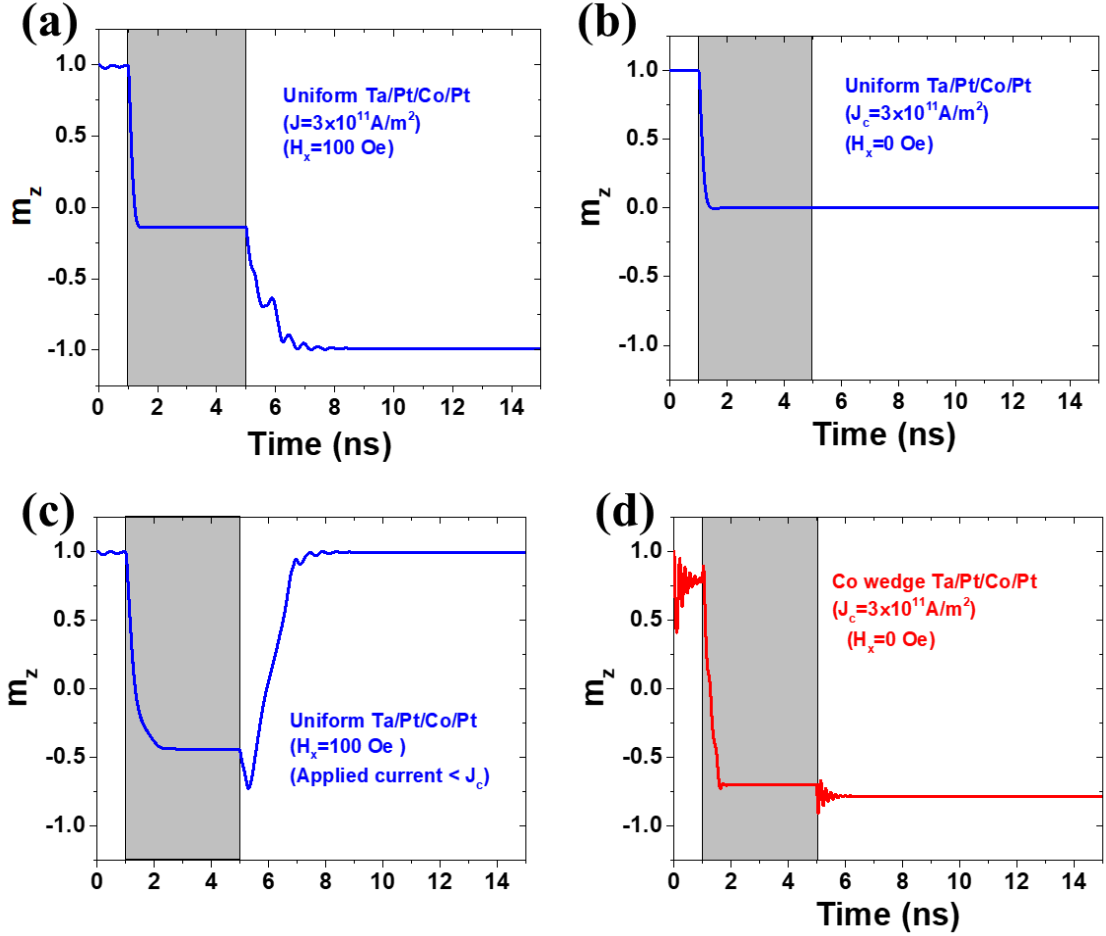

**Fig. 6(a)** Switching of $m_z$ in presence of 100 Oe field along *x*-direction. **(b)** Time evolution of $m_z$ in absence of in-plane magnetic field. **(c)** Time evolution of $m_z$ below the critical current and in presence of 100 Oe in-plane magnetic field. **(d)** $m_z$ switching with the application of current pulse in absence of the in-plane magnetic field & 5° tilt in anisotropy direction from *z*-axis. (Shaded grey region in each plot represent the 'ON' state of current pulse)

### 3.5. Thermoelectric detection of SOT switching

The thermoelectric effects in magnetic heterostructures are considered to be parasitic in nature. For instance, anomalous Nernst effect (ANE) in magnetic heterostructures can overestimate the current-induced SOT values. This is due to the same magnetization dependent symmetry of ANE and AD-SOT. Hence, they both can contribute to a 2nd Harmonic signal.

The measurement of $R_H^{2\omega}$ for Ta stack as function of out-of-plane applied field ($H_z$) resembles an AHE type of behaviour for our PMA base stack (shown in Figure 7(a) & 7(b)).

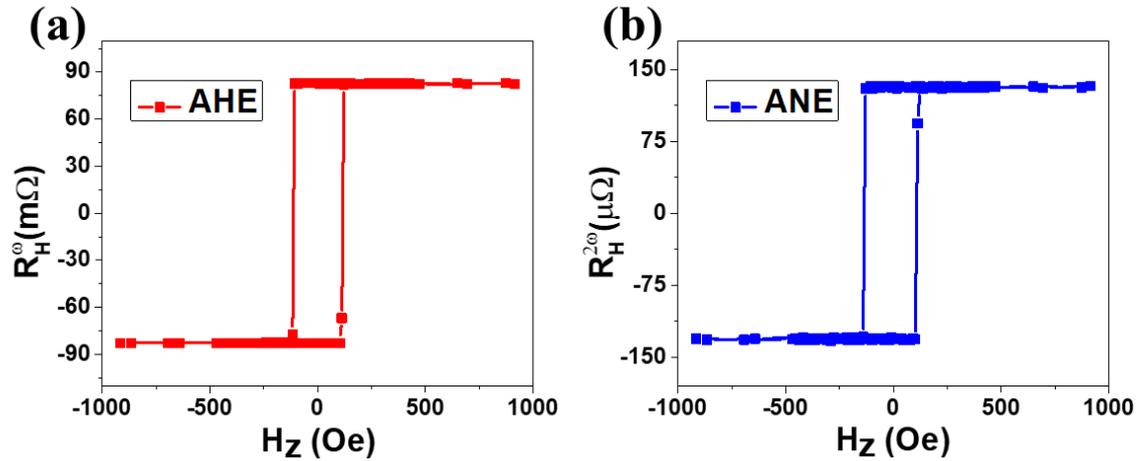

**Fig. 7(a)** Anomalous Hall resistance ($R_H$) and, **(b)** 2nd harmonic Hall resistance ($R_H^{2\omega}$) as a function of $H_z$ showing hysteric behaviour for Ta stack.

This behaviour is consistent with the ANE signature reported in the literature [28, 32]. Generally, ANE mimics anomalous Hall effect (AHE) behaviour and hence ANE signal depicts the attributes of magnetization $m_z$. Similar behaviour of AHE and ANE signifies that ANE can also be used as reading mechanism to probe different magnetization states. Conventionally for detecting the field-free magnetization switching in a PMA sample using DC current, the transverse DC resistance is measured with minute probe current after applying the varying current pulses which were swept from positive to negative magnitude and vice versa. In order to utilize ANE for detecting the magnetization orientation we have devised an AC 2nd harmonic technique. In our AC 2nd harmonic technique, a current pulse that changes the magnetization state due to AD-SOT is applied and the 2nd harmonic Hall signal instead of DC Hall resistance, is measured after the pulse. The writing and reading mechanisms are illustrated in Figure 8. At first, we have applied a 1ms current pulses with predefined magnitude to change the magnetization state utilizing AD-SOT and wait for 60 sec (similar to the DC Hall measurement scheme as described in supporting information). The second harmonic Hall resistance ($R_H^{2\omega}$) using 4 mA AC current was measured subsequently. We have taken the average of 30 readings for better signal to noise ratio.

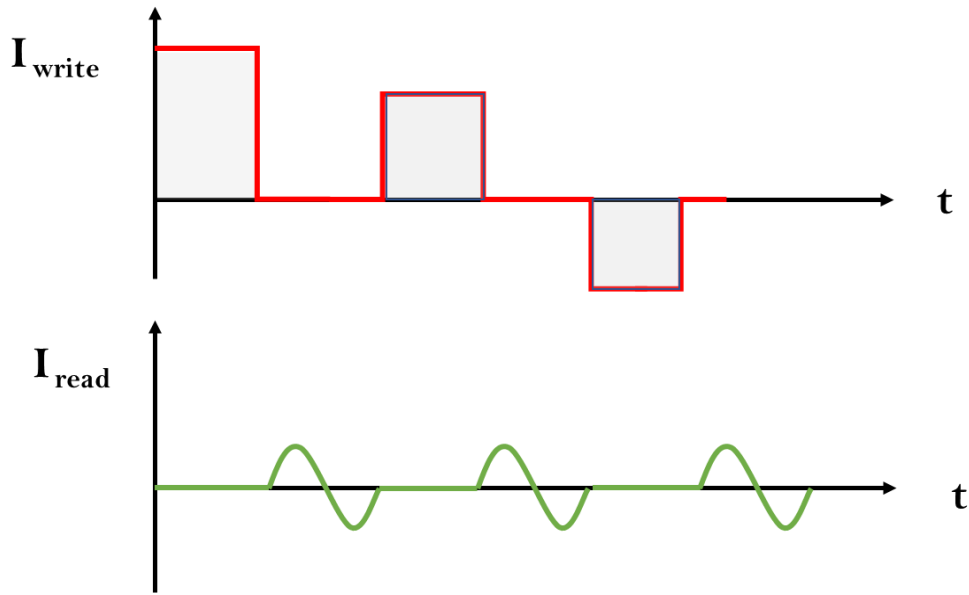

**Fig. 8** Writing and reading scheme for the 2$^{nd}$ harmonic ANE based reading mechanism. A current pulse applied to change the magnetization orientation and a subsequent AC current provided to read the 2$^{nd}$ harmonic Hall resistance of the sample.

The ANE signal in 2$^{nd}$ harmonic Hall measurement shows magnetization dependency as:

$$V^{2\omega,ANE} \propto m \times \nabla T$$

where, $\nabla T$ is the thermal gradient generated due to the Joule heating of the stack ($\nabla T \propto I^2$) and *m* is the magnetization vector of the stack. Since the 2$^{nd}$ harmonic Hall voltage is probed along *y*-direction so the above equation suggests that $V^{2\omega,ANE}$ must have a following form:

$$V^{2\omega,ANE} \propto (\nabla_x T m_z + \nabla_z T m_x)$$

Here, $\nabla_x T$ and $\nabla_z T$ are the temperature gradient along *x*- and *z*-directions and $m_x$ and $m_z$ are *x*- and *z*-component of magnetization, respectively. The $\nabla_z T$ is generated because of the differences in thermal conductivity of substrate and stack materials. Due to the different width of current and voltage branches, heat conduction occurs differently at the centre of Hall bar and towards the Hall branches, which generates $\nabla_x T$. For our high PMA based Ta stack the $m_z$ part dominates in above equation. The variation of ANE ($V^{2\omega,ANE}$) vs. $H_z$ is asymmetric in nature as the ANE signal is odd with respect to the magnetization reversal.

Further, we have employed our AC 2nd harmonic method to detect the field-free switching in Ta stack which exhibit a significant ANE contribution as depicted in Figure 3 & 7(b). Figure 9 exhibits $R_{xy}^{2\omega}$ vs. the applied current pulses for Ta stack depicting field-free magnetization switching, employing our method. As expected, $R_{xy}^{2\omega}$ is symmetric about the current axis and the complete current sweep retraces its nature with two distinct crossover regions where the magnetization switching occurs (inset of Figure 9). Using our method, we can restrict the application of the DC pulses until zero while starting from the positive (negative) value. The sensitivity of the proposed methodology was further confirmed by plotting $R_{xy}^{2\omega}$ vs the applied current pulses together with the Hall resistance (R$_H$) as a function of DC current as shown in Figure 9. The switching current estimated to be ~43 mA by our method matched well to the conventional DC Hall measurement method.

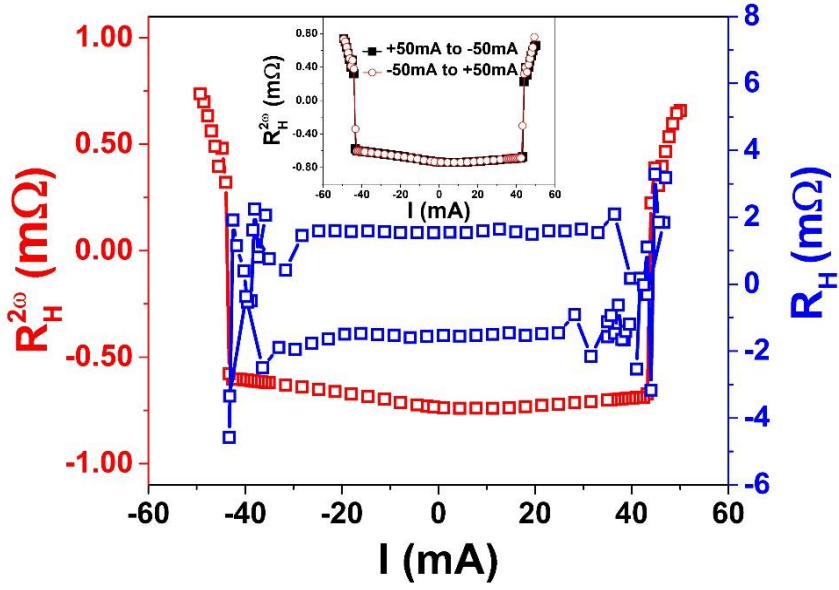

**Fig. 9** $R_H^{2\omega}$ (left *y*-axis) and $R_H$ (right *y*-axis) as a function of applied current pulses and the inset shows retracing of $R_{xy}^{2\omega}$ as a function of applied current pulses.

## 4. CONCLUSION

In summary, we have investigated the effect of addition of the Ta underlayer to the asymmetric Pt/Co/Pt stack on the enhanced perpendicular magnetic anisotropy, spin orbit torque and emergence of the ANE which was otherwise negligible in the asymmetric Pt/Co/Pt stack. It was found that FL-SOT torque remains insignificant in both the stacks whereas the AD-SOT exhibits similar magnitude. For estimating AD-SOT torque, we have adopted an angle resolved magnetization rotation technique through the measurement of $R_H^{2\omega}$ in presence of an externally applied magnetic field. Using this methodology, we could quantify the AD-SOT and ANE contributions, respectively. In addition, we could probe the field-free magnetization switching using current pulses in Ta/Pt/Co/Pt stack with significant AD-SOT. Regardless of the presence of the competing spin currents in the Ta/Pt/Co/Pt stacks, the field-free switching was attributed to the tilted anisotropy induced due to the deposition of Co in a wedge form. This was further confirmed by the micromagnetic simulation. Moreover, we had developed a methodology of detecting field-free switching by utilizing the ANE behaviour present in our sample. In our methodology, the field-free switching as well as the estimation of the switching current can be achieved by sweeping the current from zero to positive (negative) value unlike the full current sweep from positive to negative current values for the DC method. Our result suggests that the simple addition of Ta to the model Pt/Co/Pt stack leads to the unravelling of the new physical phenomena indicating that underlayer indeed plays a major role which would eventually lead to the realization of high efficiency SOT devices coupled with strong PMA.

## ASSOCIATED CONTENT

**Supporting Information:** Details of AHE and PHE contribution separation, anisotropy field ($H_k$) calculations, FL-SOT in Ta/Pt/Co/Pt stack and, current induced magnetization switching in Ta/Pt/Co/Pt stack.

## AUTHOR INFORMATION

**Corresponding Author**

**Debangsu Roy-** *Department of Physics, Indian Institute of Technology Ropar, Rupnagar 140001, India;* Email: debangsu@iitrpr.ac.in

**Authors**

**Raghvendra Posti-** *Department of Physics, Indian Institute of Technology Ropar, Rupnagar 140001, India;* orcid.org/0000-0003-1632-0095

**Abhishek Kumar-** *Department of Physics, Indian Institute of Technology Ropar, Rupnagar 140001, India*

**Dhananjay Tiwari-** Advanced safety & User Experience, Aptiv Services, Krakow 30-707, Poland

**ACKNOWLEDGMENTS**

DR acknowledges the financial support from the Department of Atomic Energy (DAE) under project no. 58/20/10/2020-BRNS/37125 & Science and Engineering Research Board (SERB) under project no. CRG/2020/005306. AK acknowledges the financial assistance from UGC. All the authors thank Prof. R G Pillay for many fruitful discussions related to this work.

# Supplementary Information

## Detection of field-free magnetization switching through thermoelectric effect in Ta/Pt/Co/Pt with significant spin-orbit torque and competing spin currents

Raghvendra Posti [1], Abhishek Kumar[1], Dhananjay Tiwari[2], and Debangsu Roy[*1]

[1]Department of Physics, Indian Institute of Technology Ropar, Rupnagar 140001, India

[2] Advanced safety & User Experience, Aptiv Services, Krakow, Poland

**S1. Separation of AHE and PHE contributions:**

Anomalous and planar Hall (AHE and PHE) contributions are separated by symmetrization and anti-symmetrization of Hall resistance data with respect to the external magnetic field. $R_H^\omega$ as a function of in-plane external magnetic field (applied along $\varphi_H \sim 45^0$) is shown in Fig.S1(a). Fig S1(a) has both AHE and PHE contributions. AHE have an odd symmetry with respect to the reversal of magnetization ($R_{AHE} \propto m_z$) while PHE have an even symmetry with respect to the magnetization reversal ($R_{PHE} \propto m_x m_y$). It is evident that the anti-symmetrization and symmetrization of $R_H^\omega$ data with respect to field axis can separate the AHE and PHE contribution, respectively [1]. Hence $R_H^\omega(-H_z \to +H_z)+ R_H^\omega(+H_z \to -H_z)$ provides AHE contribution while $R_H^\omega(-H_z \to +H_z)-R_H^\omega(+H_z \to -H_z)$ extract PHE contribution. Symmetrized and anti-symmetrized data is shown in Fig. S1(b). PHE to AHE ratio for Pt/Co/Pt stack is 0.02 and for Ta/Pt/Co/Pt stack is 0.002.

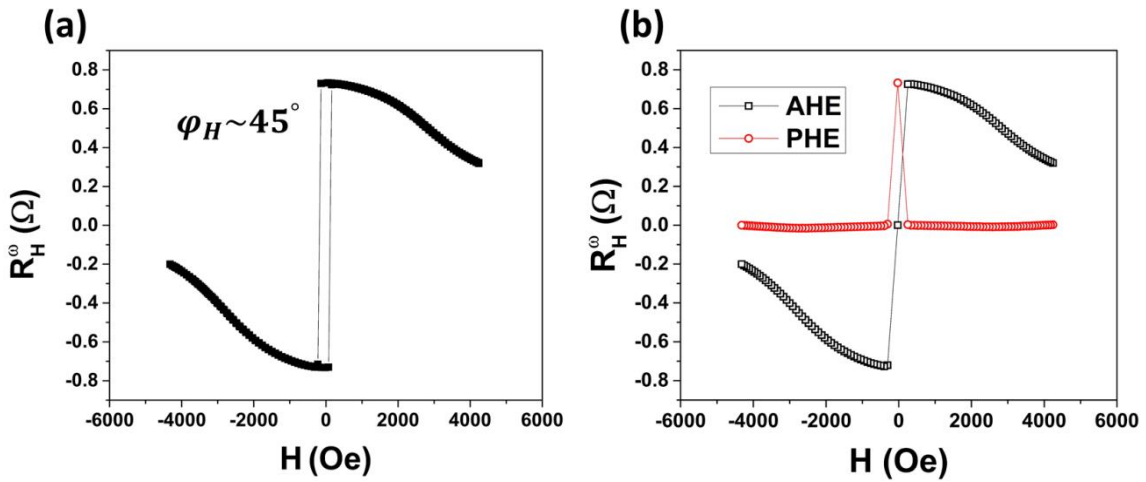

**Fig. S1 (a)** First harmonic Hall resistance ($R_H^\omega$) as a function of in-plane magnetic field applied at ~45° from current direction (For Pt/Co/Pt stack). **(b)** Separated AHE and PHE resistances as a function of in-plane field.

## S2. Anisotropy field (H$_k$) calculations:

1) **For Pt/Co/Pt sample:** First harmonic Hall voltage ($V_H^\omega$) data of Pt/Co/Pt device as a function of applied magnetic field along the current direction ($H_x$) (for coordinate geometry see Fig S2(a)) is shown in Fig. S2(b). The $V_H^\omega(H) = V_H^\omega(0)\sqrt{1 - \left(\frac{H}{H_K}\right)^2}$ fitting of this data in low field regime provides the value of anisotropy field [2] which is 3493±72 Oe.

2) **For Ta/Pt/Co/Pt sample:** According to the previous method adopted for Pt/Co/Pt stack, $H_k$ is the field value where the fitting curve cuts the *x*-axis. Ta addition enhances the anisotropy significantly rendering the switching signal very large. Thus, the curve fitting, as described for previous case, does not provide the $H_k$ value. For Ta/Pt/Co/Pt stack, we have calculated the anisotropy field using [3]:

$$sin(2\theta) = \frac{2H_{ext}}{H_k} sin(\theta_H - \theta) \qquad S(1)$$

where, $\theta_H$ is the angle between the external magnetic field ($H_{ext}$) and *z*-axis. $\theta$ is the magnetization angle (for coordinate geometry see Fig S2(a)). The method to find $\theta$ values from magnetic field rotation is discussed in the main text. The fitting of eq. S(1) is shown in Fig. S2(c). The obtained value of $H_k$ is ~1.43±0.23T for Ta/Pt/Co/Pt stack in comparison to the $H_k$ ~0.34T observed for Pt/Co/Pt stack. The enhancement in the $H_k$ due to the insertion of the Ta under layer corroborates to the literature [4].

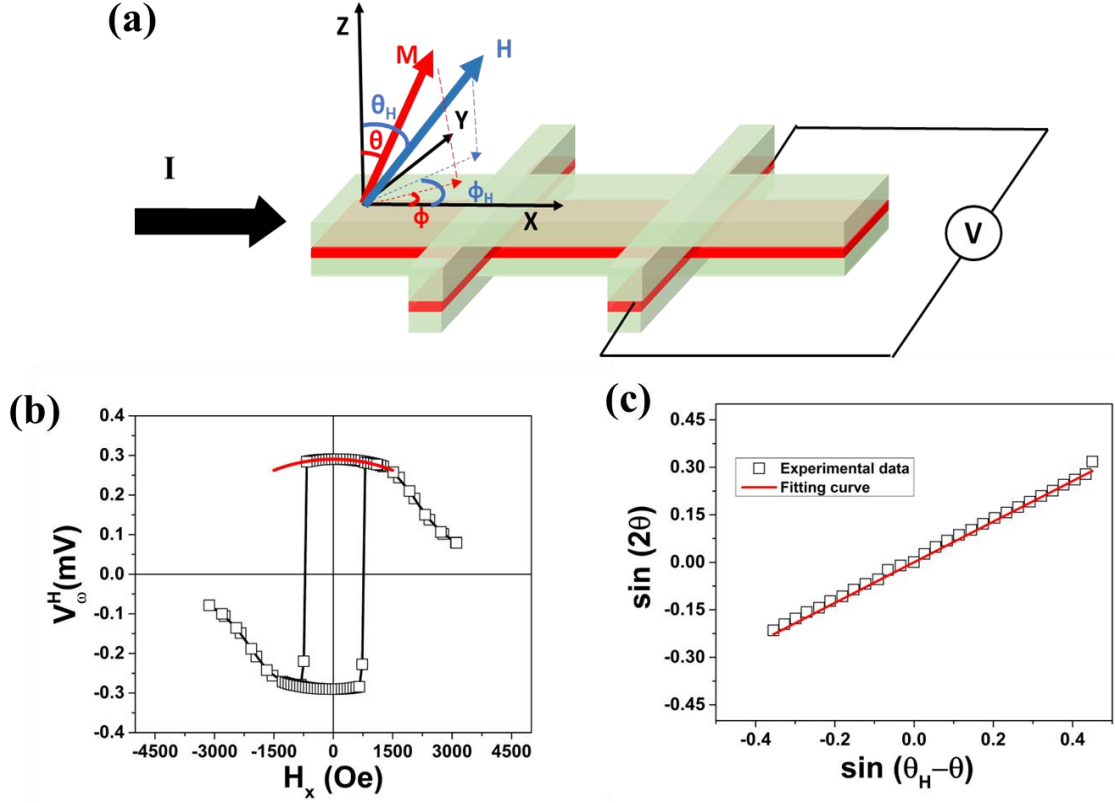

**Fig. S2(a)** Experimental geometry and coordinate system of magnetization **m**$(\theta, \varphi)$ and external magnetic field **H**$(\theta_H, \varphi_H)$. **(b)** First harmonic Hall voltage data ($V_H^{\omega}$) as a function of field along *x*-direction ($H_x$) for Pt/Co/Pt stack, the fitted curve in low field regime is shown in red color. **(c)** $sin(2\theta)$ vs. $sin(\theta_H - \theta)$ for Ta/Pt/Co/Pt stack at external magnetic field 4500 Oe.

### S3. FL-SOT for Ta/Pt/Co/Pt:

FL-SOTis determined by using the angle resolved magnetization rotation technique when $H_{ext}$ is rotated in *yz*-plane [3] (coordinate geometry described in Fig. S2(a)). The FL-SOT is induced in HM/FM heterostructures due to the interfacial Rashba-Edelstein effect (REE). For Ta/Pt/Co/Pt stack, the variation of $V_H^{2\omega}$ vs. $\theta_H$ does not show any significant trend in the resolution limit of our setup (Fig. S3). This indicates that the FL-SOT values are insignificant, similar to the Pt/Co/Pt case. The similar Pt interfaces about Co layer diminish the FL-SOT effect in both Pt/Co/Pt and Ta/Pt/Co/Pt stacks.

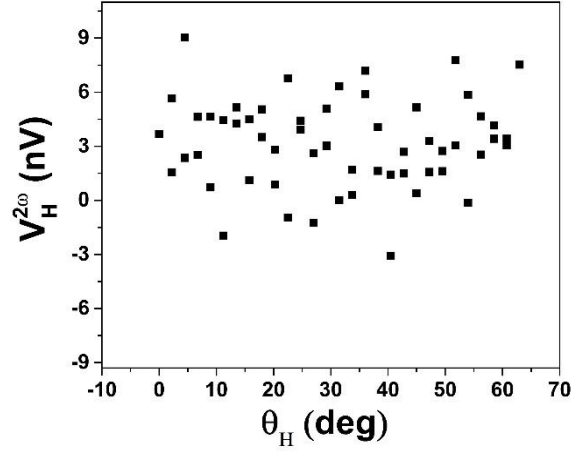

**Fig. S3.** Second harmonic Hall voltage $V_H^{2\omega}$ as a function of magnetic field angle ($\theta_H$) in *yz*-plane.

**S4. Current induced magnetization switching in Ta/Pt/Co/Pt stack:**

To verify the non-negligible value of AD-SOT, we have performed the current induced magnetization experiments in presence of the in-plane symmetry breaking magnetic fields. In this DC measurement technique, a varying current pulse for the duration of 1ms followed by a 60 sec wait time was applied through current channel. Subsequently, the transverse resistance $R_H$ was measured at 500μA read current in the Delta configuration employing Keithley 2450 source meter and 2182A nanovoltmeter. The experimental results with different in-plane field values are shown in Fig. S6. The hysteresis behaviour of $R_H$ vs. *I* plot confirms the significant value of AD-SOT in Ta/Pt/Co/Pt stack.

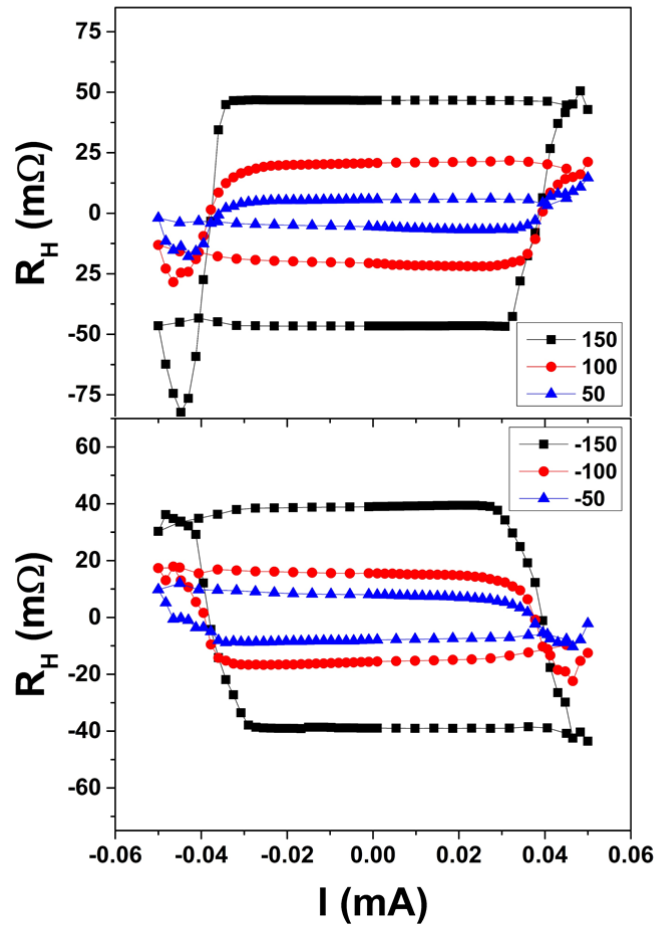

**Fig. S4.** DC Hall resistance as a function of applied current pulse in presence of in-plane applied magnetic field. (Top portion shows plots in presence of positive values of in-plane magnetic field (in Oe) and bottom portion in presence of negative values of in-plane magnetic field (in Oe)).